\documentclass[aps,prl,twocolumn,superscriptaddress,showpacs]{revtex4}

\newcommand{\beq}{\begin{equation}} 
\newcommand{\eeq}{\end{equation}} 
\newcommand{\beqn}{\begin{eqnarray}} 
\newcommand{\eeqn}{\end{eqnarray}} 
  
\usepackage{epsfig} 
\usepackage{amsmath} 
\begin{document} 
\title{Pairing fluctuations in trapped Fermi gases} 
\author{Luciano Viverit} 
\affiliation{Dipartimento di Fisica, Universit\`a di Milano, 
        Via Celoria 16, 20133, Milan, Italy} 
\affiliation{CRS-BEC INFM and Dipartimento di Fisica,
Universit\`a di Trento, I-38050 Povo, Italy}
\author{Georg M. Bruun} 
\affiliation{Niels Bohr Institute, Blegdamsvej 17, 2100 Copenhagen, Denmark} 
\author{Anna Minguzzi} 
\affiliation{NEST-INFM $\&$ Scuola Normale Superiore, Piazza dei Cavalieri 7,
        I-56126 Pisa, Italy} 
\author{Rosario Fazio} 
\affiliation{NEST-INFM $\&$ Scuola Normale Superiore, Piazza dei Cavalieri 7, 
        I-56126 Pisa, Italy} 
  
\begin{abstract} 
Fluctuations of the amplitude of the order parameter govern the
properties of superconducting systems close to the critical transition
temperature. In the BCS regime we examine the contribution of these
pairing fluctuations to the superfluid order parameter for
harmonically confined atomic Fermi gases. In the limit of small
systems we obtain an expression for the space dependence of the fluctuations,
in good agreement with  the results of numerical calculations. In this limit we
also predict a parity effect, i.e. that pairing fluctuations should
show a maximum or a minimum at the centre of the trap, depending on the value of the last 
occupied shell being even or odd. We finally propose to 
detect pairing fluctuations by measuring the density-density correlation function
as evaluated after a ballistic expansion of the gas.
\end{abstract} 
  
\maketitle 
  
Progress in trapping and cooling of gases of fermionic
atoms~\cite{fermitraps} have paved the way to explore a rich variety
of phenomena.  A major experimental effort nowadays is directed
towards the observation of the superfluid transition. The presence of
a Feshbach resonance allows to study the behaviour of Fermi gases both
in the weakly and in the strongly interacting regime by varying an
externally applied magnetic field. Away from the resonance the
effective interatomic interaction is well represented by a contact
pseudopotential $g\delta({\bf r})$, with $g=4\pi\hbar^2a/m$ and $a$
being the $s$-wave scattering length. A number of papers have
explored the properties of a Fermi system across the resonance with
particular attention to the BCS-BEC crossover and the
strongly-interacting limit $k_Fa\to \pm \infty$~\cite{strongth}, 
with $k_F=(3\pi n)^{1/3}$ and $n$ being the atomic density. In
the regime $k_Fa\to 0^+$ bosonic molecules have been already
observed~\cite{strongexps} and they have been reported to undergo
condensation~\cite{BECofmol}; current experiments start to explore the $a<0$ region \cite{exp_aneg}.
 It has been suggested that
by adiabatically tuning the scattering length from positive to
negative it could be possible to reach a BCS-like regime
\cite{carr2003}. Here we concentrate on such weak-coupling regime
$k_Fa\to 0^-$~\cite{weakth}.

Mean-field BCS theory predicts the appearence of a non-vanishing order 
parameter $\Delta_{\rm BCS}(T)$ below a critical temperature $T_c$ 
which depends strongly both on the interaction strength 
and on the geometry of 
the system~\cite{bruun}. The scale of variation of the order 
parameter is the coherence length $\xi = \hbar^2 
k_F/m\pi\Delta_{\rm BCS}(0)\gg n^{-1/3}$, $m$ being the atomic 
mass, to be compared with the size $R$ of the cloud under 
confinement (which we suppose harmonic). In the case $\xi \ll R$ of very 
large traps mean-field theory is very accurate 
in describing the physical properties of the condensate. The 
experiments with atomic gases, however, are performed on systems of 
adjustable size ranging from clouds of about $10^4-10^7$ atoms in 
harmonic traps to droplets of few tens of atoms in each lattice site 
of an optical lattice. One expects that by lowering the number of 
atoms in the trap (or equivalently when $\xi \geq R$) 
fluctuations of the order parameter can become important close to the 
critical temperature. This is the regime we discuss in this work.
Above $T_c$ fluctuations signal the onset of the 
Cooper pair instability which will eventually drive the system to the 
superfluid state.
In superconductors the role of fluctuations both on the thermodynamics 
and on the transport properties has been studied in great details both 
theoretically and experimentally and we refer to Ref.~\cite{varlamov} 
for a comprehensive review of the field. 
  
The aim of this Letter is to formulate the theory of pairing
fluctuations in 
the case of trapped Fermi gases. This problem resembles that studied 
in zero-dimensional superconductors (i.e. of radius smaller that the 
coherence length)~\cite{muhlschlegel}. There are however important 
differences that we are going to highlight. First of all pairing 
fluctuations are inhomogeneous due to the presence of the trap; we 
will show below that they are enhanced in the center of the 
trap. Another important issue is how to detect fluctuations. In 
conventional superconductors they dominate the behaviour of most 
thermodynamical and transport properties (e.g. paraconductivity, 
density of states, specific heat, $\dots$) close to 
$T_c$~\cite{varlamov}. In trapped atoms, however, those quantities are 
very hard or impossible to measure. As a clear signature for the 
detection of pairing fluctuations we propose to analyze 
density-density correlations which were recently suggested by Altman {\em et al.}~\cite{altman} as a probe of many-body states in 
ultra-cold atoms.

We assume that the system is composed of fermions in two 
internal states $\sigma=\uparrow,\downarrow$ 
with equal numbers and attractive intercomponent 
interactions, which we shall describe 
by the contact pseudopotential $g \delta({\bf r})$. 
Correspondingly, we shall consider
only pairing fluctuations in 
$s$-wave. 
This is the main collisional channel
at ultralow temperatures, while 
$s$-wave interactions among atoms belonging to the same internal
state are forbidden by the Pauli principle.
The second-quantized Hamiltonian reads
\begin{eqnarray} 
\nonumber
\hat{H}&=&\sum_{\sigma=\uparrow,\downarrow}\int d^3r\; \psi^{\dagger}_{\sigma}({\bf r}) 
\left(-\frac{\hbar^2}{2m}\nabla^2+V({\bf r})-\mu\right) 
\psi_{\sigma}({\bf r})\\ 
&+& g\int d^3r\; \psi^{\dagger}_{\uparrow}({\bf r}) 
\psi^{\dagger}_{\downarrow}({\bf r})\psi_{\downarrow}({\bf r}) 
\psi_{\uparrow}({\bf r}), 
\end{eqnarray} 
where $\psi_\sigma$ are the
fermionic field operators, $\mu$ is the chemical potential, $V({\bf
r})$ the confining potential and $g<0$.  Close to $T_c$ the most
important contribution beyond mean field comes from pair fluctuations,
we therefore treat the density correlations at the Hartree level and
add the Hartree fields $W_{\sigma}({\bf r})=g n_{-\sigma}({\bf r})$ to
the single-particle part of the Hamiltonian.

By introducing the auxiliary field $\Delta$ through a
Hubbard-Stratonovich transformation~\cite{kleinert78} the fermion
fields can be integrated over. Close to the critical temperature the
partition function becomes $Z=Z_0 \int d[\Delta^*]d[\Delta]
e^{-S[\Delta^*,\Delta]/\hbar}$, where $Z_0$ is the partition function
of the gas in the Hartree approximation and the effective action to
quartic order is
\begin{eqnarray}
\label{Seff} 
S &=& -\hbar\beta\int d{\mathbf r}_1\, d{\mathbf r}_2 
        \;\Delta^*({\mathbf r}_1)A({\mathbf r}_1,{\mathbf r}_2) 
        \Delta({\mathbf r}_2)\\ 
        &+&\frac{\hbar\beta}{2} \int \prod_{i=1,4} d{\mathbf r}_i\, 
        \Delta^*({\mathbf r}_1)\Delta^*({\mathbf r}_3) 
        B(\{{\mathbf r}_i\}) 
        \Delta({\mathbf r}_2)\Delta({\mathbf r}_4), 
\nonumber 
\end{eqnarray}
  where $\beta=1/k_BT$ and we have ignored the dependence of $\Delta$ on the 
        imaginary time
        since we are interested in the classical fluctuations above 
        $T_c$.
The kernel 
$ 
        A({\mathbf r}_1,{\mathbf r}_2)=g^{-1}\delta({\mathbf r}_1-{\mathbf r}_2) + 
        \hbar^{-1}K({\mathbf r}_1,{\mathbf r}_2) 
$ 
is defined through the two-particle propagator 
\begin{eqnarray}
\nonumber
        K({\mathbf r}_1,{\mathbf r}_2)&=& 
        -\frac{1}{2} \sum_{i,j}\frac{\tanh(\beta\xi_i/2) 
        +\tanh(\beta\xi_j/2)}{\xi_i+\xi_j}\\ 
        &\times&\phi_{i}({\bf r}_1)\phi_{j}({\bf r}_1) 
        \phi_{i}^*({\bf r}_2)\phi_{j}^*({\bf r}_2) 
\label{kernelK}
\end{eqnarray}
expressed in the Hartree-Fock basis $\{\phi_{i}({\bf r})\}$ with 
energies $\{\epsilon_i\}$ ($\xi_i=\epsilon_i-\mu$). The kernel $K$ requires 
a proper regularization as implied by the use of the contact 
pseudopotential~\cite{bruuncastin}.
The kernel of the quartic term is given by
\begin{equation} 
        B(\{{\mathbf r}_i\})=\hbar \int \prod_{i=1,3}d\tau _i G_0(1,4) 
        G_0(1,2)G_0(3,2)G_0(3,4). 
\label{b1234} 
\end{equation} 
where $G_0(a,b)$ is the single-particle Green's function and the
variable label $a=1,..,4$ stands for $({\bf r}_a,\tau_a)$. 
  
\noindent 
{\it \underline{Pairing fluctuations -}} 
In order to evaluate the pair fluctuations $\langle |\Delta({\bf
r})|^2 \rangle =Z^{-1} \int d[\Delta^*] d[\Delta] |\Delta|^2
e^{-S[\Delta^*,\Delta]/\hbar}$
 it is convenient to 
decompose $\Delta({\mathbf{r}})$ in normal modes 
\begin{equation} 
        \Delta({\bf r}) = \sum_{\nu} \chi_\nu ({\bf r}) 
        \tilde \Delta_\nu , 
\label{modeex} 
\end{equation} 
where $\chi_\nu ({\bf r})$'s are the eigenvectors of the kernel 
$A$~\cite{footnote2} 
\begin{equation} 
        \int d^3r_1\, A({\bf r},{\bf r}_1)\chi_\nu ({\bf r}_1)=\alpha_\nu 
        \chi_\nu ({\bf r}). 
\label{eigen} 
\end{equation} 
For temperatures larger than $T_c$ the dominant contribution to the 
fluctuations comes from the quadratic part of the effective action and in 
particular, from  the lowest eigenvalue 
 $\alpha_{0}$ (and the associated eigenvector 
$\chi_0 ({\bf r})$) 
\begin{equation} 
\label{chir} 
        \langle |\Delta({\bf r})|^2\rangle \sim \frac{k_BT_c}{\alpha_{0}} 
        |\chi_0 ({\bf r})|^2 \;\; . 
\end{equation} 
 For a generic system under confinement 
both $\alpha_0$ and $\chi_0({\bf r})$ have to be calculated 
numerically by diagonalization of the kernel $A$. 

For an isotropic harmonic trap $V(r) = m \omega^2
r^2/2$ we proceed now to derive analytical results in the 
small-system limit, i.e.~when
$R_{TF}\ll \xi$, where $R_{TF}=(2\epsilon_F/m\omega^{2})^{1/2}$ 
is the Thomas Fermi radius of the gas and $\xi$ the bulk BCS coherence 
length estimated at 
the center of the cloud.
Such condition 
is equivalent to $k_BT_c^{\rm bulk}\ll\hbar\omega$, where $T_c^{\rm bulk}$ 
is the bulk prediction for the critical temperature of the BCS transition 
evaluated using the central density of the cloud. In this limit the 
pairing energy is much smaller than the trap level spacing and the 
Cooper pairs are only formed between particles with quantum numbers 
$(n,l,m)$ and $(n,l,-m)$ residing in the same harmonic oscillator 
shell (``intrashell pairing regime'')~\cite{bruun}. 
Since for an isotropic trap the lowest eigenmode is spherically 
symmetric the eigenvalue 
equation (\ref{eigen})
becomes \cite{footnote3}
\begin{equation}
\label{EigvalEqIntra} 
        [1-|g|\alpha_0(T)]\chi_0(r)=\int dr'\,r'^2\; g K(r,r')\chi_0(r'), 
\end{equation} 
where the kernel describing spherically symmetric solutions is 
\begin{eqnarray} 
\nonumber 
        K(r,r')&=&-\frac{1}{8\pi}\sum_{n,n',l} 
        (2l+1) 
        \frac{\tanh(\beta\xi_{nl})+\tanh(\beta\xi_{n'l}/2)} 
        {\xi_{nl}+\xi_{n'l}}\\ 
        &\times& R_{nl}(r)R_{n'l}(r)R_{nl}(r')R_{n'l}(r'). 
\end{eqnarray} 
In the small-system limit the effect of the Hartree field is
negligible and the $R_{nl}(r)$'s are approximately the radial
eigenfunctions of the isotropic harmonic oscillator. Correspondingly
$\xi_{nl}\simeq(n+3/2)\hbar\omega-\mu$, where $\mu\simeq
\epsilon_F=(n_F+3/2)\hbar\omega$ is the chemical potential. We further
assume that pairing strictly only occurs within the same oscillator
shell, and that it mostly occurs about the Fermi surface.  In particular, we simplify
Eq.~(\ref{EigvalEqIntra}) by setting $ \int dr'\,r'^2
\;R_{nl}(r')R_{n'l}(r')\chi_0(r')\simeq \delta_{n,n'}\chi_{n} $
\cite{footnote4}, by assuming a weak dependence on $n$ for the
relevant shells around the Fermi level, and by taking
angular averages.
 The above
assumptions lead to the final expression for the normalized
eigenfunction $\chi_0(r)$ and the corresponding eigenvalue $\alpha_0$
which determine the spatial and temperature dependence of the pairing
fluctuations in Eq.~(\ref{chir}),
\begin{equation} 
\label{chi0is} 
        \chi_0(r)=\sqrt{\frac{15}{2}}\frac{(\pi a_{osc})^{3/2}}{(2
n_F+3)^{5/2}} \sum_l
\frac{(2l+1)}{4\pi}|R_{n_Fl}(r)|^2 
\end{equation}
\begin{equation}
        \alpha_0(T)= \frac{4}{15\pi^3} 
                \frac{\beta _c
R_{TF}}{a_{osc}^4}\left(1-\frac{T_c}{T}\right), 
\label{alpha0intrashell} 
\end{equation} 
where $T_c$ has been computed in Ref.~\cite{bruun},
$a_{osc}=(\hbar/m\omega)^{1/2}$ is 
the harmonic oscillator length and
we have kept only 
the leading temperature dependence in $\alpha_0$. An explicit
expression for $\chi_0(r)$ can be obtained by using that $\sum_l
(2l+1)|R_{n_Fl}(r)|^2/4\pi= \partial\rho(r)/\partial n_F$, with
$\rho(r)$ being the single-component density profile. In the Thomas-Fermi
approximation for $\rho(r)$ this yields 
$\chi_0(r)=(15/8\pi
R_{TF}^3)^{1/2} (1-r^2/R_{TF}^2)^{1/2}$, showing that 
the gas is mostly susceptibile to pairing fluctuations in the central
part of the trap.

\begin{figure}[htbp] 
  \begin{center} 
    \scalebox{.6}{\rotatebox{0}{ 
    \includegraphics[]{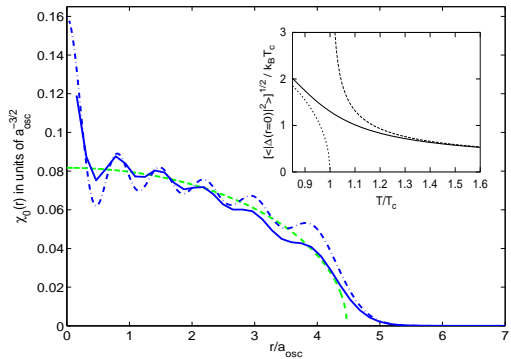}}} 
    \caption{Lowest eigenvector $\chi_0(r)$ (in units of
$a_{osc}^{-3/2}$) as a function of the radial coordinate $r$ (in units
of $a_{osc}$), as obtained from numerical diagonalization (solid line),
compared with the analytical solution (\ref{chi0is}) (dot-dashed line) and
with the Thomas-Fermi approximation (dashed line). The inset shows 
$\langle |\Delta(r=0)|^2\rangle^{1/2}$ 
(in units of $k_BT_c$) as a function of temperature (in units of $T_c$)
as obtained from the quartic (solid line) and the quadratic
approximation (dashed line) of
the effective action, and the mean-field solution (dotted line).} 
    \label{FIG1} 
  \end{center} 
\end{figure} 
  
The spatial behaviour $\chi_0(r)$
of the pairing fluctuations is illustrated in Fig.~\ref{FIG1} for a
system with $n_F=10$ filled shells 
($\sim 500$ particles) and 
coupling constant $|g|=0.7\, \hbar\omega a_{osc}^3$, as obtained from
the analytical expression (\ref{chi0is}) 
 as well as from the numerical results
obtained by diagonalization of the regularized kernel $A$ with a
procedure analogous to that described in Ref.\cite{bruuncastin}.
For our choice of parameters
$k_BT_c/\hbar\omega=0.038\ll 1$, and the system is well within the
intrashell regime ($T_c \simeq 1\,$nK). We find that the analytical
expression well approximates the numerical
solution, while the characteristic oscillating behavior due to
the discrete level structure is lost in the Thomas-Fermi
approximation. 
From the full quantum shell-structure calculations we
also predict a parity effect, i.e. that pairing fluctuations should
show a maximum or a minimum at $r=0$, depending if $n_F$ is even or odd.

\begin{figure} 
   \psfig{file=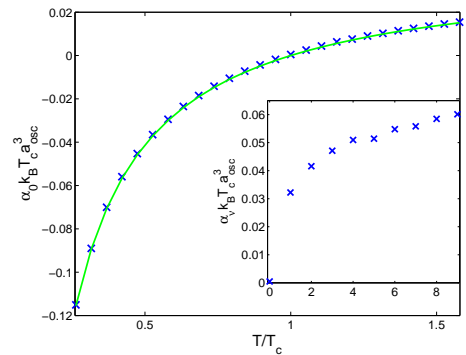,height=0.2 
       \textheight,angle=0} 
    \caption{Lowest eigenvalue $\alpha_0(T)$ in
 units of $(k_BT_c)^{-1}a_{\rm osc}^{-3}$ as a function of temperature
(in units of $T_c$) as calculated numerically
 ($\times$'s) and analytically via Eq.~(\ref{alpha0intrashell})
 (solid line). The
 inset shows the lowest eigenvalues $\alpha_\nu$ (in the same units as $\alpha_0$) as functions of the index $\nu$,
evaluated at $T\sim T_c$.}
    \label{FIG2} 
\end{figure}

At $T= T_c$ the lowest eigenvalue $\alpha_0$ vanishes, indicating the
onset of the superfluid phase. The temperature behaviour of $\alpha_0$
is illustrated in Fig.~\ref{FIG2}, where we compare the approximate
expression (\ref{alpha0intrashell}) with the exact numerical solutions
for the same choice of parameters as in Fig.~\ref{FIG1}.  The figure
shows an excellent agreement between the two calculations.

Since $\alpha_0(T_c)=0$, according to Eq.~(\ref{chir}) the pairing 
fluctuations diverge at $T_c$ (dashed line in  
the inset of Fig.~\ref{FIG1}). This is an artifact due to the quadratic
approximation used for the effective action, and can be cured by
taking into account the quartic term~\cite{muhlschlegel}, which we  estimate
from Eq.~(\ref{Seff})
within a single-mode approximation. This is well justified
by noticing that the lowest eigenvalue is significantly smaller than
the other ones (inset in Fig.~\ref{FIG2}), and thus the $\nu=0$ mode
strongly dominates the fluctuations in the intrashell regime.  The
effective action then reduces to the simple form
\begin{equation} 
\label{sefftrapfin} 
        S_{\rm eff} \simeq \hbar\beta_c\left[ 
        \alpha_0|\tilde \Delta_{\nu=0}|^2+\frac{1}{2}B_0(T_c) 
|\tilde \Delta_{\nu=0}|^4\right]. 
\end{equation} 
where $B_0(T_c)\simeq 64\beta_c^3/675\pi^6n_Fa_{osc}^6$ 
and we have only retained the contribution to the Green's functions
in Eq.~(\ref{b1234}) coming from the levels at the Fermi surface,
consistently with the intrashell pairing approximations. By
introducing the dimensionless quantity
$\Delta_{\nu=0}/(a_{osc}^{3/2}k_BT_c)$, $S_{\rm eff}$ only depends on
$n_F$ and $T/T_c$, since $R_{TF}/a_{osc}=(2n_F+3)^{1/2}$. The value of 
$n_F$ controls the magnitude of the critical region around $T_c$, 
which we estimate to be $\delta T/T_c \sim 2/(n_F+1)(n_F+2)$ from the Ginzburg criterion. 

Using the effective action Eq.~(\ref{sefftrapfin}) the
temperature behaviour of the pairing fluctuations can be estimated
analytically, and is illustrated in the inset of Fig.~\ref{FIG1}
(solid line).  The figure shows that the 
divergence predicted by the quadratic term alone is indeed eliminated 
and the behavior is now smooth across $T_c$, tending to the
mean-field result (dotted line) below $T_c$. 
We should remark that Eq.~(\ref{sefftrapfin})
is only accurate close enough to $T_c$, while well below $T_c$ the full 
action should be withheld in the averaging. 

We also mention that in the large-system limit $\xi
\ll R_{TF}$ pairing fluctuations can be described along the lines 
of Ref.~\cite{barpetr}. We do not pursue the issue further in this
work since in this limit they are expected to be small.

\noindent 
{\it \underline{Density correlations - }} The presence of Cooper pairs
implies non-zero density-density correlations among particles with
opposite spins located at ${\bf q}$, $-{\bf q}$. Altman {\em et
al.}~\cite{altman} suggested that such ``anomalous'' correlations
could be measured probing the density-density correlation function
after the trap has been released and the gas has undergone a ballistic
expansion: $\langle n_\sigma({\bf r}) n_{-\sigma}({\bf r}')\rangle
\sim \langle n_\sigma({\bf q}_{\bf r}) n_{-\sigma}({\bf q}_{\bf
r'})\rangle$, where ${\bf q}_{\bf r} = (m/\hbar t) {\bf r} $, $m$ is
the mass of the fermions and $t$ is the time lapse after the trap has
been opened. At mean-field level it can be shown that $\langle \delta
n_\sigma({\bf r}) \delta n_{-\sigma}(-{\bf r})\rangle
\propto|\Delta_{BCS}|^2$, with $\delta n_\sigma({\bf r})=
n_\sigma({\bf r})-\langle n_\sigma({\bf r}) \rangle $, and hence, one
could directly access the gap field amplitude. Above $T_c$ the same
density-density correlation function is also sensitive to pairing
fluctuations. Proceeding as for obtaining Eq.~(\ref{Seff}) for a
uniform system we get $\langle \delta n_\sigma({\bf r}) \delta
n_{-\sigma}(-{\bf r})\rangle = \langle \langle (\Delta/2E_{q_r})^2
[1-2 f(E_{q_r})]^2 \rangle \rangle _{S_{eff}} \;$, where $\langle
\langle \cdot \rangle \rangle _{S_{eff}}$ means the average over the
distribution of $\Delta$ weighted by the effective action $S_{eff}$,
$f(x)=\{\exp[\beta(x-\mu)]+1\}^{-1}$ is the Fermi distribution and
$E_{q}=(\hbar^4q^4/4m^2+\Delta^2)^{1/2}$ is the usual BCS quasi-particle
dispersion relation.

In small harmonically trapped systems anomalous correlations occur
between particles with opposite spins and quantum numbers $(n,l,m)$
and $(n,l,-m)$. 
In the intrashell limit analogous
approximations to those used in order to obtain Eqs. (\ref{chi0is})
and (\ref{alpha0intrashell}) yield
\begin{equation}
\langle \delta n_\sigma({\bf r}) \delta n_{-\sigma}(-{\bf r})\rangle=
\frac{1}{4}\left(\frac{\partial\rho(q_r)}{\partial n_F}\right)^2 
\langle\langle [F(\Delta)]^2\rangle\rangle_{S_{eff}}, 
\end{equation} 
where $F(\Delta)=1-2f(|\Delta|)+2(|\Delta|/\hbar\omega)\log(\gamma
n_F)$, with $\gamma=e^C\simeq 1.78$, and $C$ Euler's
constant. In this limit the spatial dependence of the anomalous
correlations is proportional to $|\chi_0(r)|^2$, which is given by
Eq.~(\ref{chi0is}) and shown in Fig.~\ref{FIG1}. Also for the small
system the mean-field result is non-zero only below $T_c$, where we
find $\langle \delta n_\sigma({\bf r}) \delta n_{-\sigma}(-{\bf r})\rangle
\propto
|\Delta_{BCS}({\bf r})|^2$.

The effect of pairing fluctuations is not confined to the density-density 
correlations described in this work. As another example, we expect a soft gap 
in  the laser induced tunneling (discussed in Ref.~\cite{torma})
above the critical temperature.

The authors would like to thank Prof. M. Tosi for helpful discussions. 
This work has been supported by the EU (RTN-Nanoscale). LV wishes to thank 
the SNS, where part of this work was carried out, for the kind
hospitality. AM wishes to thank the LPTMS Orsay, where this work was
completed, for the kind hospitality.


\begin{thebibliography}{30}
\bibitem{footnote3} In the derivation of the analytical results in
	place of the regularization of the
	kernel we use a sum cutoff.
 	Such procedure is justified a posteriori by comparison with the
	numerical results based on the proper renormalization scheme.
\bibitem{footnote4}
	This follows from the oscillating nature of $R_{nl}(r)$ combined 
	with the fact that $\chi_0(r)$ has a definite sign.
\end{thebibliography}
\end{document}